\documentclass[conference]{IEEEtran}
\IEEEoverridecommandlockouts
\usepackage{placeins}
\usepackage{orcidlink}
\usepackage{subcaption}
\usepackage{cite}
\usepackage{amsmath,amssymb,amsfonts}
\usepackage{algorithmic}
\usepackage{graphicx}
\usepackage{textcomp}
\usepackage{xcolor}
\usepackage{booktabs}
\usepackage{multirow}
\usepackage{array}
\usepackage{url}
\usepackage{adjustbox}
\usepackage{comment}
\usepackage{pifont}

\newcommand{\cmark}{\ding{51}} % ✓
\newcommand{\xmark}{\ding{55}} % ✗

\usepackage{fancyhdr}
\fancypagestyle{plain}{
  \fancyhf{}
  \fancyhead[L]{\footnotesize 5th IEEE International Conference on AI in Cybersecurity (ICAIC), 18-20 February 2026, University of Houston, Houston, United States}
  \fancyfoot[C]{\thepage}

}

\makeatletter
\let\ps@IEEEtitlepagestyle\ps@plain
\makeatother

\begin{document}
\title{AgentSOC: A Multi-Layer Agentic AI Framework for Security Operations Automation\\
\thanks{\vspace{-1cm}979-8-3315-4970-1/26/\$31.00~\copyright~2026 IEEE}
}
\author{\IEEEauthorblockN{1\textsuperscript{st} Joyjit Roy\,
%\orcidlink{0009-0000-0886-782X}
}
\IEEEauthorblockA{\textit{IEEE Member} \\
Austin, Texas, USA \\
joyjit.roy.tech@gmail.com}
\and
\IEEEauthorblockN{2\textsuperscript{nd} Samaresh Kumar Singh\, %\orcidlink{0009-0008-1351-0719}
}
\IEEEauthorblockA{\textit{IEEE Senior Member} \\
Leander, Texas, USA \\
ssam3003@gmail.com}
}

\maketitle

\begin{abstract}
Security Operations Centers (SOCs) increasingly encounter difficulties in correlating heterogeneous alerts, interpreting multi-stage attack progressions, and selecting safe and effective response actions. This study introduces AgentSOC, a multi-layered agentic AI framework that enhances SOC automation by integrating perception, anticipatory reasoning, and risk-based action planning. The proposed architecture consolidates several layers of abstraction to provide a single operational loop to support normalizing alerts, enriching context, generating hypotheses, validating structural feasibility, and executing policy-compliant responses. Conceptually evaluated within a large enterprise environment, AgentSOC improves triage consistency, anticipates attackers' intentions, and provides recommended containment options that are both operationally feasible and well-balanced between security efficacy and operational impact. The results suggest that hybrid agentic reasoning has the potential to serve as a foundation for developing adaptive, safer SOC automation in large enterprises. Additionally, a minimal Proof-Of-Concept (POC) demonstration using LANL authentication data demonstrated the feasibility of the proposed architecture.
\end{abstract}

\begin{IEEEkeywords}
Agentic AI, Security Operations Center, Threat Detection, MITRE ATT\&CK, Counterfactual Reasoning, Structural Simulation
\end{IEEEkeywords}
%==============================================================================
% I. INTRODUCTION
%==============================================================================
\section{Introduction}
\label{sec:introduction}

\textbf{Security Operations Centers (SOCs)} encounter increasing complexity within contemporary enterprise environments. Analysts contend with high alert volumes, fragmented security tools, and continuously evolving attacker tactics. According to surveys, enterprise SOCs receive over 100,000 alerts per day\cite{alert_fatigue_2024}. As such, analysts spent between 2 and 4 hours on manual triage, with nearly 70\% of these alerts being left without further investigation ~\cite{khayat2025empowering, nagar2018evolution}. This resulted in alert fatigue, inconsistent decisions, and delayed response.

Existing SIEM, SOAR, and XDR platforms automate certain capabilities, but still rely significantly on human interpretation. These systems correlate events and initiate predefined workflows. However, analysts remain ultimately accountable for contextual decision-making and executing final response actions.

Large Language Models (LLMs) have introduced advanced capabilities in summarization, contextual enrichment, and rule-driven automation \cite{srinivas2025ai}. However, most of these LLM solutions act as "copilots". They suggest actions but rely on human investigation. They also lack grounding in enterprise constraints and may recommend responses that are unsafe or infeasible within existing identity and network structures \cite{alert_fatigue_2024}. As a result, creative generation and operational safety remain misaligned in current solutions.

A significant gap persists. SOCs require autonomous reasoning capable of anticipating attacker movement, validating structural feasibility, and selecting response actions that safeguard business operations. Such autonomy must be explainable, risk-aware, and aligned with policy controls to avoid accidental disruption. Existing tools do not provide these capabilities.

This paper presents \textbf{AgentSOC}, a multi-layer agentic AI framework designed for SOC environments. AgentSOC is intended to support autonomous decision-making with context awareness and safety alignment. The framework integrates perception, anticipatory reasoning, and risk-aware action planning within a unified operational loop.

AgentSOC addresses the limitations of current SOC automation through three integrated mechanisms:

\begin{enumerate}
    \item \textbf{Narrative Counterfactual Engine (NCE)}: Generates multiple attack hypotheses using LLM-based reasoning grounded in MITRE ATT\&CK tactics, techniques, and procedures. This mechanism enables anticipatory threat detection beyond traditional pattern matching.
    \item \textbf{Structural Simulation Engine (SSE)}: Validates NCE-generated hypotheses against enterprise topology, identity privilege graphs, and policy constraints. This process filters infeasible attack paths and ensures structural realism.
    \item \textbf{Risk Scoring and Evaluation Module (RSEM)}: Ranks defensive actions using composite scores that balance containment effectiveness, business impact, and execution cost. This module supports autonomous and operationally safe response decisions.
\end{enumerate}

The primary contributions of this paper are as follows:

\begin{itemize}
    \item A multi-layer autonomous SOC framework that integrates perception, agentic reasoning, and policy-constrained action execution.
    \item Risk-aware action selection through RSEM, which quantifies security and business tradeoffs.
    \item Multi-hypothesis attack reasoning through NCE, grounded in MITRE ATT\&CK knowledge.
    \item Graph-based feasibility validation using SSE to ensure structural realism.
    \item A proof-of-concept (POC) demonstration utilizing LANL authentication data achieved sub-second processing times (506 ms), thereby demonstrating the feasibility of the integrated reasoning approach~\ref{sec:evaluation}).
\end{itemize}

AgentSOC is evaluated using a conceptual enterprise scenario. The results demonstrate improved triage coherence, anticipatory detection, and safer containment strategies. This approach has the potential to reduce analyst workload and enhance enterprise defense posture.

%===================================================================
% II. BACKGROUND AND RELATED WORK
%===================================================================

\section{Background and Related Work}
\label{sec:related}

\subsection{SOC Automation Challenges}
Traditional SIEM platforms aggregate security logs and apply rule-based 
correlation, but struggle with multi-stage attack detection and generate 
excessive false positives, with research showing that more than 70\% of alerts represent false positives or low-priority events\cite{khayat2025empowering}
\cite{alert_fatigue_2024}\cite{shah2022managing}. SOAR systems automate predefined playbooks but lack adaptive reasoning, executing static workflows without evaluating the consequences or business impact of actions \cite{nagar2018evolution}. XDR platforms improve visibility but do not unify context across identity, topology, and privilege structures. As a result, the analyst workload remains unsustainable. The global cybersecurity workforce shortage, estimated at 3.5 million unfilled positions \cite{isc2_workforce_2023}, intensifies these challenges. Studies indicate that security teams allocate more time to alert triage than to actual threat investigation \cite{patel2023automated}.

\subsection{LLM Assisted SOC Operations}
Recent work explores LLMs for threat intelligence summarization, alert enrichment, and analyst assistance \cite{srinivas2025ai}. Commercial systems such as Microsoft Security Copilot provide natural language interfaces but function as advisory tools\cite{srinivas2025ai}. They suggest responses but do not make autonomous decisions or evaluate feasibility within enterprise constraints \cite{alert_fatigue_2024}. Academic studies on LLM-driven attack graphs generate relationships but do not integrate them with real-time response planning or structural validation \cite{dif2025towards}. These approaches risk hallucinating infeasible attack paths and unsafe containment recommendations. They lack sufficient grounding in network topology, identity and privilege models, and policy controls.

\subsection{Attack Modeling and Knowledge Bases}
The MITRE ATT\&CK framework provides a systematic classification of attacker behaviors and methods, facilitating organized threat assessment and TTP-based detection development \cite{uralov2025using}. However, most systems use ATT\&CK for reactive rule creation rather than for generative reasoning about potential attack progressions. Attack graph modeling provides path-based analysis but typically operates on static snapshots without integration into live SOC workflows \cite{sarraf2022deepdefender}. Existing graph models rarely evaluate whether a recommended action is safe for production environments. They also rarely include business impact or execution risk in decision-making.

\subsection{Agentic AI and Autonomous Cyber Defense}
Counterfactual reasoning, defined as the ability to explore alternative futures and compare action outcomes, has been applied in planning and robotics. However, it remains underutilized in security operations~\cite{pearl2018why}. Agentic AI research in cybersecurity emphasizes planner-based agents, goal-directed reasoning, and autonomous task execution \cite{kshetri2025transforming}. Some systems investigate closed-loop defensive automation \cite{stefanov2025autonomous}, while others apply LLMs to explain complex alerts or generate mitigation strategies. However, no existing approach combines multi-hypothesis generation, topology-grounded feasibility checks, and quantitative risk scoring for action selection. Current methods either lack autonomy, structural grounding, or integration with SOC-scale infrastructure.

\subsection{Research Gap}
\label{subsec:research_gap}
Table~\ref{tab:capability_comparison} compares AgentSOC with existing 
SOC automation approaches across key functional dimensions.

Prior research does not present a system that combines: \\
(1) generative hypothesis creation using LLMs, \\
(2) structural validation against enterprise topology and identity privilege graphs, \\
(3) quantitative risk-aware action scoring that considers business continuity, and \\
(4) closed-loop autonomous operation with outcome-aware adjustment.

AgentSOC integrates these four capabilities into a unified framework. It performs anticipatory reasoning, validates decisions against enterprise constraints, and selects actions that defend effectively while preserving operational continuity and safety \cite{kshetri2025transforming}\cite{stefanov2025autonomous}.

\begin{table*}[t]
\centering
\caption{Comparison of AgentSOC with Existing SOC Automation Approaches}
\label{tab:capability_comparison}
\small
\begin{adjustbox}{width=\textwidth}
\begin{tabular}{@{}lllll@{}}
\toprule
\textbf{Capability} & \textbf{Manual SOC} & \textbf{SIEM/SOAR} & \textbf{LLM Copilot} & \textbf{AgentSOC} \\
\midrule
Alert Synthesis & Manual correlation & Rule-based clustering & LLM summarization & Autonomous enrichment \\
Attack Path Reasoning & Analyst judgment & Limited correlation & NL inference (unvalidated) & Multi-hypothesis counterfactual \\
Structural Validation & Manual queries & None & None & Graph-based feasibility \\
Response Selection & Analyst discretion & Fixed playbooks & Suggested actions & Risk-optimized ranking \\
Business Impact Awareness & Stakeholder consult & Hard-coded thresholds & Policy-aware recommendations & Real-time parameter integration \\
Autonomous Closed-Loop & No & Limited (playbook) & No (analyst-in-loop) & Yes (perception$\rightarrow$reasoning$\rightarrow$action) \\
\bottomrule
\end{tabular}
\end{adjustbox}
\end{table*}

%===================================================================
% III. SYSTEM ARCHITECTURE
%===================================================================

\section{System Architecture}
\label{sec:architecture}

AgentSOC is a multi-layer agentic AI framework that integrates perception, anticipatory reasoning, and risk-aware action planning within a closed-loop operational cycle. The architecture comprises four primary components:

\begin{itemize}
    \item \textbf{Perception Layer}: Ingests and enriches security alerts into structured incident objects
    \item \textbf{Agentic Reasoning Layer}: Generates and validates attack hypotheses using LLM-based counterfactual reasoning
    \item \textbf{Action and Playbook Layer}: Plans and executes policy-compliant defensive responses
    \item \textbf{Internal Knowledge Store}: Maintains enterprise context, topology, and execution state
\end{itemize}

This architecture supports closed-loop autonomous operation. Alerts move through a pipeline where they are normalized, enriched with enterprise metadata, analyzed with counterfactual reasoning, validated against structural constraints, scored for risk-aware response selection, and finally converted into policy-compliant response plans. Our current proof-of-concept completes this cycle in under one second (approximately 506ms).Figure~\ref{fig:architecture} shows the end-to-end workflow.

\textbf{Scope and Assumptions.} It has been assumed that the adversaries can perform credential theft, lateral movement, and ATT\&CK-aligned privilege escalation. Zero-day detection and malware analysis are outside the current scope. The system processes SIEM and EDR signals and operates within enterprise IAM and network visibility boundaries.

%================================================================
%PERCEPTION LAYER
%================================================================
\subsection{Perception Layer}
The Perception Layer serves as the sensory component of AgentSOC. Its role is to convert heterogeneous security signals into structured incident objects for downstream reasoning. It comprises 3 core subcomponents:

\textbf{Alert Normalization:} Ingests alerts from SIEM platforms, EDR tools, NDR appliances, cloud security controls, and OS-level log sources. Data are transformed into a unified schema maintaining uniform fields for timestamps, severity levels, and event sequencing. In practice, this enables cross-source correlation that would otherwise require manual work.

\textbf{Situational Contextualization:} Adds metadata such as asset profiles, identity attributes, privilege paths, and network topology from the Internal Knowledge Store. Each alert becomes an incident object enriched with the context needed for hypothesis generation (for example, user role, authentication strength, privilege tier, policy compliance, and risk indicators).

\textbf{Noise Reduction:} Deduplicates and clusters related alerts, enriches incidents, and filters out low-confidence or repetitive signals while keeping notable events. The intent is to reduce analyst workload without weakening coverage.

%================================================================
%AGENTIC REASONING LAYER
%================================================================
\subsection{Agentic Reasoning Layer}

The Agentic Reasoning Layer introduces the system’s core reasoning capability by integrating hybrid predictive reasoning. It combines generative LLM-based hypothesis formation with graph-based structural validation and quantitative risk evaluation. This module has the following 3 subcomponents.

\textbf{Narrative Counterfactual Engine (NCE):} Uses LLMs to propose multiple plausible attack progressions for each enriched incident. When detecting anomalous authentication activity, the module produces multiple prospective attack paths, including credential exploitation, horizontal network traversal, and Kerberos-driven elevation of privileges. Each scenario includes a confidence score and an explanation of the underlying evidence. The engine also highlights missing context and anchors its outputs to MITRE ATT\&CK techniques \cite{mitre_attack_2024}.

\textbf{Structural Simulation Engine (SSE):} Checks whether NCE-generated scenarios are possible within the real environment. It traverses the identity and privilege graph to determine if the required network paths, privilege transitions, and ATT\&CK technique preconditions are satisfied. Scenarios that violate structural constraints are discarded, while conditionally feasible ones are retained with clear dependency notes.

\textbf{Risk Scoring and Evaluation Module (RSEM):} Ranks defensive actions using a weighted score that balances containment value against business impact:
\begin{equation}
\text{Composite Score} = (\alpha \times \text{Containment}) - (\beta \times \text{Business Impact})
\end{equation}

where $\alpha$ and $\beta$ are tunable weights that reflect organizational risk tolerance. This scoring approach prioritizes actions that maximize threat mitigation while minimizing service disruption and compliance violations.

%===================================================================
% VI. ACTION AND PLAYBOOK LAYER
%===================================================================
\subsection{Action and Playbook Layer}

The Action and Playbook Layer converts reasoning outputs into operational response steps that can be executed reliably within the enterprise environment. This Layer includes 3 subcomponents:

\textbf{Adaptive Playbook Generator:} Builds multi-step workflows by pairing feasible attack paths with the defensive actions ranked by the reasoning layer. Action primitives include \texttt{REVOKE\_SESSION}, \texttt{RESTRICT\_PRIVILEGES}, \texttt{ENABLE\_MFA}, \texttt{QUARANTINE\_ACCESS}, and \texttt{MONITOR\_ONLY}. These are combined based on context rather than a fixed template.

\textbf{Policy and Safety Guardrails:} Before any workflow is approved, it is checked against business-impact thresholds, operational dependencies, and compliance requirements. Playbooks with high projected impact are routed to analysts; lower-risk ones move forward to autonomous execution.

\textbf{Execution Interface:} Connects to SOAR platforms, EDR agents, and IAM systems to carry out approved actions. Executions start in dry-run mode, generating audit logs and offering rollback paths for safety. Full execution is enabled only when the environment conditions meet the required constraints.

%=============================================================
%SUPPORTING COMPONENTS
%=============================================================
\subsection{Supporting Components}
Apart from the 3 core components, the architecture relies on following 2 supporting components that maintain enterprise context, validate post-execution outcomes, and maintain the closed-loop feedback cycle essential for safe autonomous operation:

\textbf{Internal Knowledge Store:} A unified repository that holds enterprise context and operational state. It integrates asset and service metadata (CMDB), identity–privilege graphs (IAM), network topology mappings (SDN controllers), MITRE ATT\&CK transitions, business-impact parameters (GRC platforms), and compliance constraints into a structured knowledge base. This consolidated view defines the boundaries and guardrails for autonomous responses, ensuring they remain architecturally feasible and policy compliant. Alongside this static context, the store maintains dynamic behavioral knowledge: derived threat indicators, historical traces, telemetry deltas, and post-execution state updates from perception and action cycles. These signals keep predictive reasoning aligned with the current environment rather than a fixed snapshot.

\textbf{Real-Time Monitoring:} Observes the environment during and after response execution. It captures host and network state changes, correlates new alerts with recent decisions, and checks whether containment actions produced the intended results. Execution summaries, deviation flags, and rollback indicators feed back into the Knowledge Store, forming a closed loop. This reduces drift between planned and actual behavior and supports safe, incremental improvement of automated decision-making.

%=============================================================
%END TO END WORKFLOW
%=============================================================
\subsection{Operational Workflow}

AgentSOC functions through a continuous \textbf{Sense–Reason–Act} cycle that integrates all architectural layers. The workflow includes the following steps: 
\\(1) alerts are normalized and enriched with enterprise context, 
\\(2) NCE generates multiple attack hypotheses, 
\\(3) SSE validates structural feasibility, 
\\(4) RSEM ranks defensive actions, 
\\(5) Playbooks are synthesized and evaluated against policies, 
\\(6) Approved actions are executed, 
\\(7) Real-time monitoring captures outcomes, and 
\\(8) the Knowledge Store is updated for subsequent cycles.

\begin{figure*}[htbp]
\centerline{\includegraphics[width=0.9\textwidth]{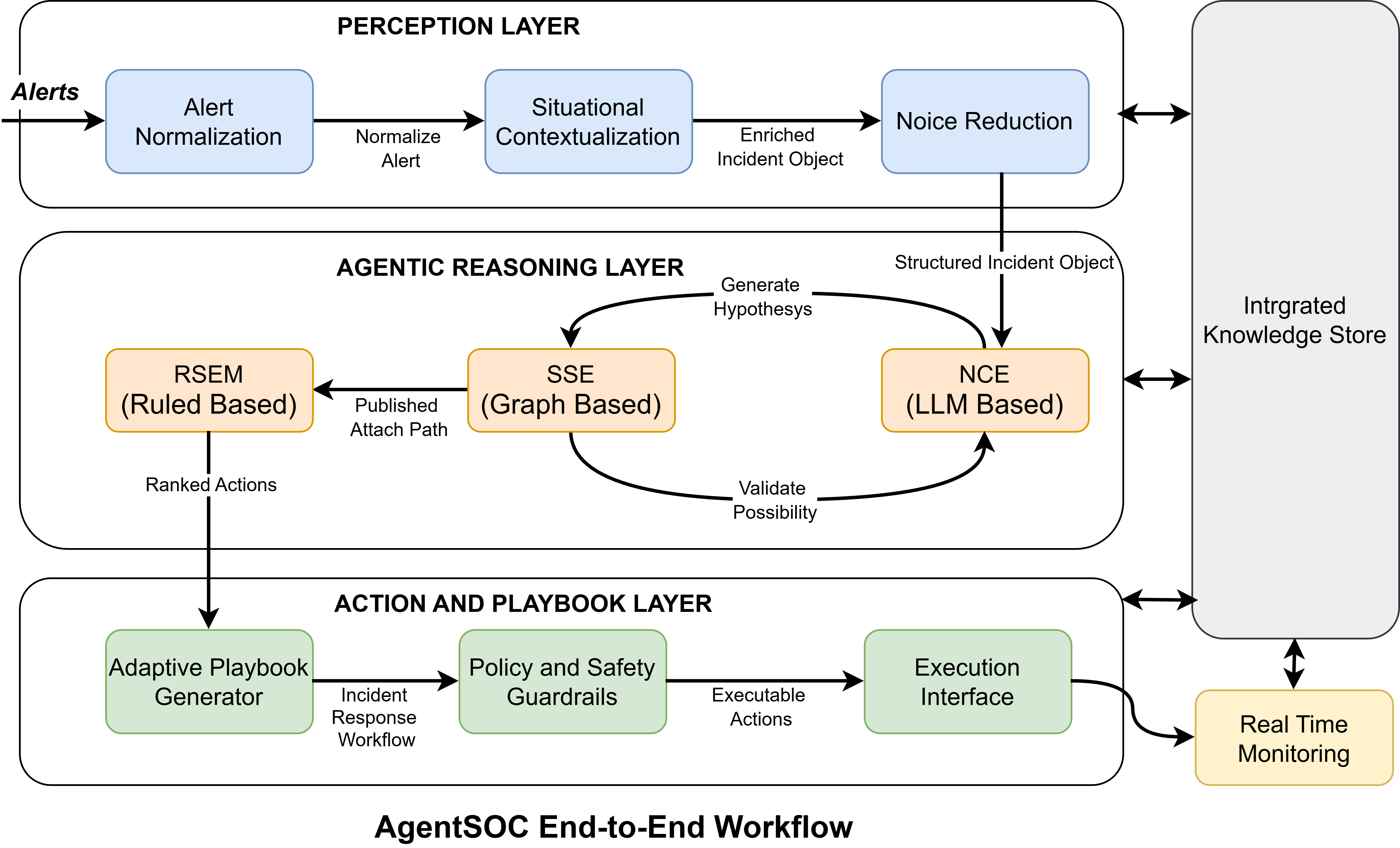}}
\caption{End-to-end AgentSOC workflow shows the continuous \textbf{Sense-Reason-Act cycle}. Alerts flow through Perception, Agentic Reasoning, and Action layers. The Internal Knowledge Store provides enterprise context, and Real-Time Monitoring enables closed-loop feedback.}
\label{fig:architecture}
\end{figure*}

This closed loop enables anticipatory reasoning. The system reacts to current events, forecasts where an attacker may go next, and tests response effectiveness before acting to keep operations autonomous and safe. Figure~\ref{fig:architecture} illustrates the end-to-end operational workflow.

%===================================================================
% IV. PROOF-OF-CONCEPT EVALUATION
%===================================================================

\section{Proof-of-Concept Evaluation}
\label{sec:evaluation}

\subsection{Evaluation Setup}
\label{subsec:setup}

To validate the core ideas beyond the architecture diagrams, a small proof-of-concept (POC) was developed using a 5,000 event sample from the \textbf{LANL Comprehensive, Multi-Source Cyber-Security Events dataset}. The subset contains Kerberos authentication logs from the Los Alamos National Laboratory network, which provides real enterprise signals such as routine logins and occasional credential misuse patterns.

\textbf{Environment:} Local workstation (Intel i7-9700K, 32~GB RAM) running Python~3.10. The prototype relies on NetworkX for graph construction and NumPy/Pandas for event handling. GPT-4 was used only to propose candidate attack hypotheses for comparison. The enterprise graph was modeled as a synthetic 50 node topology to approximate host reachability and privilege flows.

\textbf{Focus of the POC:} The goal was not to produce a deployable system. Instead, the prototype is scoped to exercise the reasoning loop: alert intake, graph enrichment, and hypothesis evaluation. Alerts were derived from authentication anomalies such as cross-domain access, repeated failures, suspicious geolocation changes, and short-interval lateral moves across multiple hosts.

\subsection{Proof-of-Concept Demonstration}
\label{subsec:poc}

To demonstrate the integrated reasoning pipeline, we evaluated AgentSOC using an authentication anomaly derived from the LANL dataset. The test case involves a finance analyst whose credentials are phished and used to log in to a corporate workstation (Host A). Over the next several minutes, the system produces separate alerts for suspicious PowerShell use, irregular Kerberos authentication, indicators of credential dumping, and a lateral move toward a production database server (Host B) that the analyst has never accessed before.

In conventional SIEM/SOAR workflows, these detections are generated as discrete alerts that require manual correlation by analysts \cite{nagar2018evolution}. An analyst has to confirm host relationships, verify the user's privileges, review past activity, and decide on containment. Based on industry SOC survey data, that level of triage usually takes several minutes and delays response.
\\(1) Alert Normalization and Contextual Enrichment, 
\\(2) Narrative Counterfactual Engine via LLM, 
\\(3) Structural Simulation Engine using enterprise graph, and 
\\(4) Risk Scoring and Evaluation Module for action ranking.
\\
\textbf{Example Input Alert:} 

\begin{verbatim}
SourceUser: user123
SourceHost: ws-fin-27
DestinationHost: srv-fin-03
EventType: Kerberos TGT Request
Result: Success
Timestamp: 2023-11-14 13:22:41
\end{verbatim}

\textbf{Enriched Incident Object:} Table~\ref{tab:enriched} presents the Perception Layer output after contextual enrichment.

\begin{table}[h]
\centering
\caption{Enriched Incident Object}
\label{tab:enriched}
\small
\begin{tabular}{@{}ll@{}}
\toprule
\textbf{Field} & \textbf{Value} \\
\midrule
Incident ID & INC-POC-001 \\
User & user123 (Privilege Tier 2 - Finance) \\
Source Host & ws-fin-27 (Finance workstation) \\
Target Host & srv-fin-03 (Finance DB, criticality 9/10) \\
Historical Baseline & No prior access to srv-fin-03 \\
Event Type & Kerberos TGT Request (Success) \\
Flags & unusual-TGT-request, cross-tier-access \\
\bottomrule
\end{tabular}
\end{table}

The cross-tier-access flag indicates potential privilege escalation or lateral movement, elevating the incident's risk profile.

\textbf{NCE-Generated Hypotheses:} Table~\ref{tab:hypotheses} presents the LLM-generated attack progressions. The confidence scores indicate how strongly each progression is supported by alert attributes, behavioral deviations, and ATT\&CK pattern alignment.

\begin{table}[h]
\centering
\caption{NCE-Generated Hypotheses}
\label{tab:hypotheses}
\small
\begin{tabular}{@{}clc@{}}
\toprule
\textbf{ID} & \textbf{Description} & \textbf{Confidence} \\
\midrule
H1 & Credential misuse $\rightarrow$ lateral movement & 0.74 \\
H2 & Kerberos ticket abuse $\rightarrow$ privilege escalation & 0.52 \\
H3 & Benign misconfiguration & 0.21 \\
\bottomrule
\end{tabular}
\end{table}

\textbf{H1} is identified as the most probable malicious scenario, based on the unusual cross-tier access and the absence of prior occurrences. In contrast, \textbf{H3} considers a benign explanation. However, confidence in this hypothesis (\textbf{H3}) remains low due to insufficient supporting evidence.

\textbf{Structural Feasibility (SSE).} Table~\ref{tab:feasibility} presents the graph-based validation results on the 50-node enterprise topology.

\begin{table}[h]
\centering
\caption{SSE Structural Feasibility Validation: Graph-based validation results on the 50-node enterprise topology}
\label{tab:feasibility}
\small
\begin{tabular}{@{}l l p{5.1cm}@{}}
\toprule
\textbf{Hypothesis} & \textbf{Feasible?} & \textbf{Reason} \\
\midrule
H1 & \cmark & Network path exists; group allows SMB pivot \\
H2 & \cmark~(cond.) & Feasible if Tier-1 creds exist on target \\
H3 & \xmark & No service/task associated with user123 \\
\bottomrule
\end{tabular}
\end{table}

The SSE validated \textbf{H1} by confirming network connectivity and verifying that group memberships permit SMB access. \textbf{H3} was rejected due to insufficient structural support. Only \textbf{H1} and conditional \textbf{H2} advanced to the risk scoring phase.

\textbf{Risk Scoring (RSEM):} Table~\ref{tab:rsem} depicts the action ranking using Composite Score = $(0.7 \times \text{Containment}) - (0.3 \times \text{Business Impact})$.

\begin{table}[h]
\centering
\caption{RSEM Action Ranking}
\label{tab:rsem}
\small
\begin{tabular}{@{}clccc@{}}
\toprule
\textbf{Rank} & \textbf{Action} & \textbf{Contain.} & \textbf{Impact} & \textbf{Score} \\
\midrule
1 & A1 - Isolate ws-fin-27 & 0.92 & 0.15 & 0.599 \\
2 & A2 - Disable user123 & 0.84 & 0.30 & 0.498 \\
3 & A3 - Monitor events & 0.15 & 0.00 & 0.105 \\
\bottomrule
\end{tabular}
\end{table}

Host isolation yields the highest composite score, offering robust containment (0.92) with minimal operational disruption (0.15). Monitoring alone is inadequate due to high-confidence threat indicators. \textbf{Recommendation:} Isolate ws-fin-27.

\textbf{Processing Performance:} Table~\ref{tab:processing} presents the timing breakdown demonstrating sub-second latency.

\begin{table}[h]
\centering
\caption{Processing Time Breakdown}
\label{tab:processing}
\small
\begin{tabular}{@{}lcl@{}}
\toprule
\textbf{Stage} & \textbf{Time (ms)} & \textbf{Description} \\
\midrule
Normalization & 6 & Schema transformation \\
Enrichment & 12 & Graph query for context \\
NCE (LLM) & 480 & Counterfactual Hypothesis generation \\
SSE & 8 & Graph traversal validation \\
RSEM & $<$1 & Vectorized risk calculation \\
\midrule
\textbf{Total} & \textbf{$\sim$506} & \textbf{Sub-second latency} \\
\bottomrule
\end{tabular}
\end{table}

\begin{figure}[h]
\centering
\includegraphics[width=0.35\textwidth, height=0.25\textheight]{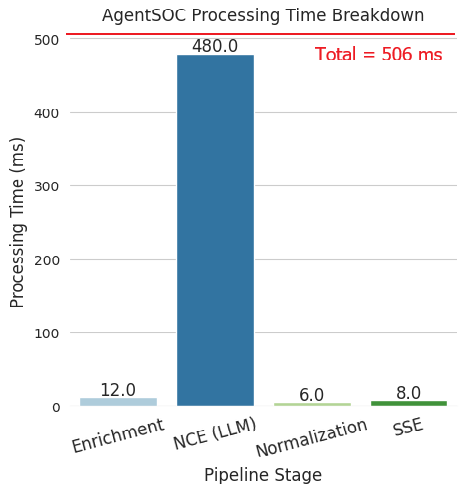}
\caption{Processing time breakdown across AgentSOC pipeline stages demonstrating sub-second latency (506 ms total).}
\label{fig:processing}
\end{figure}

The results in Figure~\ref{fig:processing} show that LLM-driven hypothesis generation consumes roughly 95\% of the total processing time. In practice, this latency is still workable for SOC operations, where most incidents develop over several minutes. The graph validation and risk scoring steps finish quickly in comparison, which suggests that this part of the design is not a bottleneck for deployment.

\subsection{Comparative Analysis}
\label{subsec:comparative}

The POC illustrates how AgentSOC changes day-to-day SOC workflows rather than replacing them outright. Table~\ref{tab:capability_comparison} (Section~\ref{subsec:research_gap}) positions AgentSOC in relation to current SOC automation approaches. It advances beyond existing systems in the following areas:

\textbf{Reduction in Triage Effort:}
Authentication events are consolidated into a single incident record containing user identity, authorization tier, asset importance, and behavioral indicators. 
In a traditional workflow, analysts correlate these items across SIEM, EDR, and IAM consoles. AgentSOC performs this step automatically (18 ms in the POC), which reduces manual lookups without removing analyst oversight.

\textbf{Autonomous closed-loop operation:}
SIEM/SOAR platforms offer limited, playbook-based automation, and LLM copilots generally operate with an analyst in the loop. AgentSOC executes the perception–reasoning–action cycle under policy constraints and can complete it without manual intervention.

\textbf{Anticipatory Reasoning:}
Conventional tools react to what has already happened and rely on predefined detections. AgentSOC's NCE module attempts to forecast likely attacker actions such as lateral movement or privilege escalation. This is a tentative capability rather than a guarantee, but it did surface patterns earlier than rule-based matching in the POC.

\textbf{Structural validation:}
Current tools do not combine LLM hypothesis generation with graph-based feasibility checks. As a result, they are vulnerable to infeasible or hallucinated paths. LLM assistants can propose attack paths that do not make sense in the environment. AgentSOC uses Structural Soundness Evaluation (SSE) to filter these proposals against topology and privilege boundaries. In the POC, hypothesis H3 was rejected because the required path did not exist, which reduces false leads and errors.

\textbf{Risk-aware action selection:}
Traditional responses depend on fixed playbooks or analyst judgment. Automation platforms often act only on technical impact. AgentSOC ranks containment options by combining technical urgency with business constraints, such as downtime tolerance, and by estimated containment effectiveness and business impact, rather than treating all actions as equal. This does not remove the need for human approval, but it offers a starting point that aligns with policy.

\textbf{Latency and Real-Time Use:}
The reasoning loop completed in approximately 506 ms. End-to-end timing stayed under one second across repeated runs. This suggests the design is light enough for real-time use in SOC environments, even with LLM components.

Together, these capabilities support an approach that combines generative reasoning, structural grounding, and policy-aware automation for adaptive SOC operations.
%===================================================================
% VIII. DISCUSSION
%===================================================================
\section{Discussion}

\textbf{Operational Benefits:} 
The proof-of-concept (POC) shows that AgentSOC can support SOC operations by combining alert normalization, contextual enrichment, and anticipatory reasoning in a single workflow. Heterogeneous detections are synthesized into enriched incident objects, and the NCE module proposes likely attack progressions. In practice, this reduces the amount of manual triage and improves situational awareness earlier in the incident lifecycle.

\textbf{Structural Validation:} 
SSE validates hypothetical attack paths against network topology and privilege boundaries before they are used for reasoning. This step constrains LLM-generated hypotheses and limits interpretations that do not match the environment. The approach differs from unconstrained LLM assistants~\cite{srinivas2025ai}, which do not verify their outputs against organizational structure or policy.

\textbf{Risk-Aware Action Selection:}
RSEM incorporates policy constraints and business impact parameters when ranking containment options. This helps AgentSOC select responses that are effective without introducing unnecessary operational risk. The approach targets a limitation in static SOAR playbooks~\cite{nagar2018evolution}, which struggle to adjust actions based on changing enterprise conditions or to distinguish between high- and low-impact interventions. Within AgentSOC, quantitative scoring supports alignment with organizational requirements rather than applying a single containment strategy in all cases.

\textbf{Hybrid Architecture Advantages:}
The hybrid agentic architecture combines generative reasoning, structural validation, and risk-aware automation. LLMs support hypothesis generation and explanation in natural language, while graph-based validation filters infeasible paths. Policy-aware scoring ranks the remaining options for safety and cost. Together, these components address gaps in current SOC automation. Rule-based systems are rigid, LLM assistants lack environmental grounding, and static playbooks provide limited context awareness.

\textbf{Practical Deployment Considerations:}
In the POC, LLM latency averaged 480 ms, which is within SOC tolerance since incident timelines are measured in minutes~\cite{khayat2025empowering}. Maintaining the Knowledge Store is essential, as topology and privilege data must remain current to ensure accurate feasibility validation. The conservative escalation approach, in which high-risk actions require human approval, builds analyst trust prior to full automation. Dry-run mode with comprehensive audit logging provides accountability and rollback capability essential for enterprise adoption

%===================================================================
% IX. LIMITATIONS AND FUTURE WORK
%===================================================================
\section{Limitations and Future Work}
\label{sec:limitations}

\textbf{Current Limitations:} Testing was conducted using a fixed sample from the LANL authentication dataset~\cite{akent2015lanl}, instead of continuous production telemetry. This limits what can be said about real-time adaptation. The enterprise graph includes only 50 nodes, which simplifies the complexity found in large production estates. MITRE ATT\&CK mappings~\cite{mitre_attack_2024} are rule-based, so they do not capture zero-day tactics or fast-moving adversary behavior. Response actions were tested in dry-run mode and did not produce closed-loop feedback from operational systems. 
Business impact scores use simplified heuristics instead of live operational metrics. Observed LLM latency (480 ms) may also need optimization to sustain higher event volumes in practice.

\textbf{Future Directions:}
A primary step is the integration of live telemetry streams to support continuous operation. Future efforts aim to explore reinforcement learning for adaptive risk scoring based on execution outcomes~\cite{stefanov2025autonomous}. Further work includes developing production-grade SOAR connectors with rollback paths and adding human-in-the-loop checkpoints to improve reliability and adoption. Extending the design to multi-cloud and hybrid IT/OT environments (AWS, Azure, GCP)~\cite{alert_fatigue_2024} is an active area of interest. Another direction is integrating threat intelligence feeds to adjust defensive posture proactively rather than reacting only to observed activity.

\begin{comment}
\textbf{Current Limitations.} The evaluation uses a static dataset (LANL authentication logs~\cite{akent2015lanl}) instead of live telemetry, which constrains the validation of real-time adaptation. Using a 50-node enterprise graph reduces the complexity of actual production networks. MITRE ATT\&CK mappings~\cite{mitre_attack_2024} are rule-based and may not adequately represent zero-day tactics or adaptive adversary behaviors. Response actions were evaluated in dry-run mode, lacking closed-loop feedback from production environments. Business impact assessments are based on simplified heuristics rather than dynamically measured operational metrics. The observed LLM latency (480ms) may require further optimization to support high-volume operational settings.

\textbf{Future Directions.} Key extensions include integration of live telemetry streams to enable continuous operation, and employing reinforcement learning for adaptive risk scoring based on execution outcomes~\cite{stefanov2025autonomous}. Additional priorities include developing production SOAR interfaces with rollback capabilities, implementing human-in-the-loop feedback mechanisms to enhance model reliability and foster organizational trust, extending applicability to multi-cloud (AWS, Azure, GCP) and hybrid IT/OT environments ~\cite{alert_fatigue_2024}, and integrating with threat intelligence platforms to facilitate proactive adjustment of the defense posture.
\end{comment}
\section{Conclusion}
\label{sec:conclusion}

This paper presented \textbf{AgentSOC}, a multi-layer agentic architecture designed to support Security Operations Center (SOC) workflows. The system integrates perception, anticipatory reasoning, and risk-aware action selection. The framework incorporates three primary mechanisms, (1) the Narrative Counterfactual Engine for LLM-based hypothesis generation, (2) the Structural Simulation Engine for graph-constrained feasibility checks, and (3) the Risk Scoring and Evaluation Module for policy-aware response ranking. Together, these components aim to combine the flexibility of generative reasoning with the reliability of structure-constrained validation.

A proof-of-concept (POC) implementation using the LANL authentication dataset~\cite{akent2015lanl} achieved sub-second runtime (506 ms). In this evaluation, AgentSOC reduced manual triage effort through automated enrichment and supported earlier reasoning about attacker intent than rule-based correlation alone. Structural validation filtered out infeasible LLM outputs, and policy-aware scoring helped align response choices with operational requirements.

These results suggest that AgentSOC can serve as a basis for more autonomous SOC workflows while remaining within enterprise safety boundaries.By integrating perception, reasoning, and action in a continuous sense-reason-act cycle, the architecture advances beyond current SIEM/SOAR platforms and LLM-assisted tools, moving toward fully autonomous, adaptive, and explainable security operations.

\bibliographystyle{IEEEtran}
\bibliography{references}

\end{document}